\documentclass[twocolumn,english,aps,pre, eqsecnum]{revtex4-2}
\usepackage[T1]{fontenc}
\usepackage[latin9]{inputenc}
\usepackage{bm}
\usepackage{amsmath}
\usepackage{graphicx}
\usepackage{geometry}
\usepackage{esint}

\makeatletter

\providecommand{\tabularnewline}{\\}

\makeatother

\usepackage{babel}
\usepackage{listings}

\begin{document}
\title{Complexity order of multiple resource algorithms}
\author{Run Yan Teh, Manushan Thenabadu, Peter D Drummond}
\address{Centre for Quantum Science and Technology Theory, Swinburne University
of Technology, Melbourne 3122, Australia}
\email{peterddrummond@protonmail.com}

\begin{abstract}
Algorithmic efficiency is essential to reducing energy and time usage
for computational problems. Optimizing efficiency is important for
tasks involving multiple resources, for example in stochastic calculations
where the size of the random ensemble competes with the time-step.
We define the complexity order of an algorithm needing multiple resources
as the exponent of inverse total error with respect to the total resources
used. The optimum order is predicted for independent, factorable resources.
We show that it equals the inverse sum of the inverse resource orders.
This is applied to computing averages in a stochastic differential
equation. We treat numerical examples for multiple different algorithms
and for stochastic partial differential equations, all giving quantitative
results in excellent agreement with our more general analytic theory.
\end{abstract}
\maketitle

\section{Introduction}

Computational complexity is important to computer science, but is
also of increasing significance in other disciplines that use computers,
including physics, mathematics, and operations research. One can quantify
computational complexity in terms of the resources required to solve
a particular computational problem \citep{hartmanis1971overview}.
Other measures of computational complexity include communication complexity
\citep{comm_complexity_quant_switch,PhysRevResearch.6.043269}, circuit
complexity \citep{Cir_Complexity} and parallel computing complexity
\citep{compl_parallel,parallel_comp}. Here, we analyze the optimization
of resource use for continuous algorithms with multiple resources
and errors. These challenges arise in interdisciplinary problems,
and especially in simulating emergent and collective phenomena in
physics. We demonstrate a general optimization method that enhances
algorithmic efficiency. 

Resource optimization plays a crucial role in solving computational
problems, since the resources required depend on the complexity of
the problem and the algorithm. The time and energy for these tasks
can significantly add to the economic cost of the solutions. Optimizing
energy consumption was a precursor to the development of quantum computing
\citep{landauer1961irreversibility}, and carries ecological benefits.
Reducing energy use in computational data centers is important, since
they now contribute significantly to global carbon emissions \citep{masanet2020recalibrating}.

We particularly focus on stochastic methods whose computational applications,
originally in physics \citep{Karatzas1991brownian,VanKampen2007stochastic,Gardiner2009Stochastic,klebaner2012introduction,Drummond2014Quantum,Arnold1992stochastic},
now extend to interdisciplinary applications in biology, chemistry,
engineering, medicine, complex systems and quantum technologies. Such
methods involve averages over ensembles of random numbers. They have
the lowest errors in the limit of large ensembles, but computational
results always use finite ensembles.

 For ordinary differential equation solvers, minimizing complexity
gives algorithms of high time-step order, allowing solutions of a
given error with fewer steps, less time and lower energy. For many
interdisciplinary problems, reducing the time-step error competes
with other resources. The complexity order is introduced here to describe
the resulting order after optimization. We ask: what complexity order
is obtained most generally, with multiple error sources?

To resolve this question, we investigate computational resources with
multiple independent errors. We find that the optimum complexity order
equals the inverse sum of the inverse individual orders. This is a
general property of factorable, independent resources. We evaluate
this for several cases, and give numerical examples. One of many applications
of understanding complexity is the question of determining if there
is a quantum advantage when solving problems on a quantum computer.
Such tasks are often stochastic \citep{arute2019quantum}, and comparisons
should make use of the optimal classical approach.

Stochastic differential, partial differential and stochastic partial
differential equations all require multiple resources. As an example,
we apply complexity optimization to stochastic differential equations.
One can improve the step-size error or improve the sampling error,
but these compete for resources. Using a fixed resource of either
one is not convergent. Both must be varied simultaneously to obtain
convergence, and a fixed ratio is not optimal. The resulting complexity
order $c$ has the range $1/3\le c<1/2$.

More generally, we find the optimal scaling exponent for multiple
resources. Numerical examples are given with more than one problem
and numerical algorithm. They are applied to stochastic and stochastic
partial differential equations. These illustrate two and three resource
cases. These results were obtained on a public stochastic library,
xSPDE4 \citep{Drummond2024xSPDE4,Kiesewetter2016xSPDE,kiesewetter2023xspde3,kiesewetter2023codebase}.
Outputs were checked against independent codes to ensure reliability,
giving excellent agreement with predictions.

\section{Complexity order\label{sec:Complexity-order}}

Continuous algorithms often have multiple errors and use multiple
resources. Examples include stochastic differential equations, partial
differential equations and Monte Carlo algorithms, which are relevant
to many calculations in physics and elsewhere.

Before performing any numerical simulations, we are confronted with
the problem of deciding the number of samples, time points, and space
points to be used. We term these multiple resources for a simulation
and their finiteness results in distinct error types. To analyze such
numerical problems, we consider a general algorithm which requires
multiple resources, defined as the total number of floating point
operations used. We assume for simplicity that the resource requirements
are factorable and the error contributions are independent, in a quantitative
sense defined below.

The general definition of the convergence of a sequence of approximations
$\bm{z}(N)$ to an exact result $\bm{e}$ is that there is convergence
to an order $n$ if, for a resource $N,$ (typically inverse to the
step-size) one has that, for $m$ multiple evaluations $z_{k}$ requiring
a total resource $N$: 
\begin{equation}
\epsilon_{k}(N)=\left|z_{k}(N)-e_{k}\right|<\epsilon_{0}N^{-n}.
\end{equation}

In this paper we use a simpler definition, reducing these multiple
criteria to one, by averaging over the errors. We define the error
as the RMS or more generally the p-norm average error over a set of
$m$ evaluations, typically at multiple time and/or space points.
These could also be observations of multiple different averages: 
\begin{equation}
\epsilon(N)=\left[\frac{1}{m}\sum_{k}\left|z_{k}(N)-e_{k}\right|^{p}\right]^{1/p}.
\end{equation}

We define a resource as factorable if the total resource usage factorizes
as $N\equiv N_{A}\tilde{N}=N_{A}\prod N_{i}$, where $N_{A}$ is the
minimal 'one-step' method complexity, and $\tilde{N}$ is the number
of uses of the method. Here, we define $N_{i}$ as the resource required
to give an expected error of $\epsilon_{i}$ for each component $i$
of the algorithm. Regarding $\bm{\epsilon}=\left[\epsilon_{1},\epsilon_{2},\epsilon_{3},\ldots\right]$
as a d-dimensional vector, the errors are defined to combine independently
as a vector p-norm, if they give a total error $\epsilon$ such that
\begin{equation}
\epsilon\left(\bm{N}\right)=\left[\sum_{i=1}^{d}\epsilon_{i}^{p}\left(N_{i}\right)\right]^{1/p}.\label{eq:total error}
\end{equation}
The power $p$ depends on the type of criterion used, either an error
bound, with $p=1$, or a root mean square (RMS) average, with $p=2$.
If we assume that each independent error scales only as a power of
$N_{i}$, then $\epsilon_{i}\left(N_{i}\right)=\epsilon_{0i}N_{i}^{-n_{i}},$
where $n_{i}$ is the order of the $i$-th error. Our numerical examples
use RMS or Euclidean norm averages, but our main results are independent
of the error norm.

Examples of this include partial differential equations that propagate
in time and space, where time and space complexity factorize, and
ordinary stochastic differential equations where time and ensemble
resources factorize. For partial stochastic differential equations,
all three resources, that is, time, space and ensemble size, often
may factorize simultaneously. We now wish to evaluate the total complexity
order for such cases, defined here as the optimum exponent such that:
\begin{equation}
c=-N\frac{\partial}{\partial N}\ln\epsilon\left(\bm{N}\right).
\end{equation}
We will consider two possible optimization scenarios, of minimizing
the error at fixed resource or minimizing the resource at fixed error,
and show that these give an identical complexity exponent. We will
assume for simplicity that the errors are power laws of each $N_{i}$,
which is typically the case in an asymptotic limit, but we note that
in general the error exponents may scale dynamically at finite resources,
and give different scaling exponents depending on the resources allocated.

\subsection{Minimizing the error at fixed resource}

What is the complexity limit if one has a fixed resource, and wishes
to minimize errors? Optimizing the resource cost can be treated with
the use of Lagrange multipliers. To minimize the error with a constrained
total resource $N$, where $N\equiv N_{A}\prod N_{i}$, we define:
\begin{equation}
\epsilon\left(\bm{N},\lambda\right)=\epsilon\left(\bm{N}\right)+\lambda\left(N_{A}\prod_{i=1}^{d}N_{i}-\hat{N}\right),\label{eq:Lagrangian}
\end{equation}
where $\lambda$ is a Lagrange multiplier and $N$ is constrained
to $\hat{N}$. Differentiating with respect to each $N_{i}$, the
minimum error requires that: $\partial\epsilon\left(\bm{N},\lambda\right)/\partial N_{i}=0.$This
is obtained when the following relationship holds for all error sources,
\begin{align}
N\lambda & =n_{i}\epsilon_{i}\left(N_{i}\right)\epsilon\left(\bm{N}\right)^{1-p}=n_{i}\epsilon_{0i}^{p}\epsilon\left(\bm{N}\right)^{1-p}N_{i}^{-pn_{i}},\label{eq:Lagrange_solution}
\end{align}
 and the individual resources required for this are $N_{i}=\left[n_{i}\epsilon_{0i}^{p}\epsilon\left(\bm{N}\right)^{1-p}/(N\lambda)\right]^{1/(pn_{i})}$.
Taking a product over the $d$ solutions obtained from Eq (\ref{eq:Lagrange_solution})
gives the result that $\left(N\lambda\right)^{1/(pc)}=N_{A}N^{-1}\prod_{i}\left(n_{i}\epsilon_{0i}^{p}\epsilon\left(\bm{N}\right)^{1-p}\right)^{1/(pn_{i})}$,
where the optimal exponent $c$ is given by the central result of
this paper:
\begin{equation}
c=\left[\sum n_{i}^{-1}\right]^{-1}.\label{eq:optimal complexity order}
\end{equation}
As a result, on solving for the Lagrange multiplier $\lambda,$ one
obtains
\begin{equation}
\lambda=\frac{1}{N}\left(N^{-1}N_{A}\right)^{cp}\prod_{i}\left(n_{i}\epsilon_{0i}^{p}\epsilon\left(\bm{N}\right)^{1-p}\right)^{c/n_{i}}.\label{eq:Lagrange_multiplier}
\end{equation}
 Combining the Lagrange multiplier result with the total error definition,
Eq. (\ref{eq:total error}) and the resource solution, for $N_{i}$
gives the surprisingly elegant central result of our paper, which
is that the total error at optimum resource usage also has a power
law scaling, with:
\begin{equation}
\epsilon\left(\bm{N}\right)=\epsilon_{0}N^{-c}.\label{eq:optimum error}
\end{equation}
The prefactor $\epsilon_{0}$ and the optimum resource $N_{i}$ are
given respectively by:
\begin{align}
\epsilon_{0} & =N_{A}^{c}c^{-1/p}\prod_{i}\left(n_{i}\epsilon_{0i}^{p}\right)^{c/(pn_{i})}\label{eq:complexity_prefactor}\\
N_{i}^{n_{i}} & =N^{c}\frac{\epsilon_{0i}n_{i}^{1/p}}{\epsilon_{0}c^{1/p}}.\nonumber 
\end{align}

\subsection{Two resource case}

For the case of two resources, one immediately obtains that:
\begin{equation}
\frac{N_{1}^{n_{1}}}{N_{2}^{n_{2}}}=\frac{\epsilon_{01}}{\epsilon_{02}}\left(\frac{n_{1}}{n_{2}}\right)^{1/p}.\label{eq:resource power ratio}
\end{equation}
Since $N_{2}=\tilde{N}/N_{1}$, there is an optimum ratio of the two
resources, $r=N_{1}/N_{2}$ , given by:
\begin{equation}
r=\left[\left(\frac{\epsilon_{01}n_{1}^{1/p}}{\epsilon_{02}n_{2}^{1/p}}\right)^{2}\tilde{N}^{n_{2}-n_{1}}\right]^{1/(n_{1}+n_{2})}.\label{eq:variational_ratio}
\end{equation}
This implies that the resources allocated depend on $N$, and hence
on the target error, through Eq (\ref{eq:optimum error}). In summary,
the complexity order $c$ is given by the inverse sum of the inverse
orders $n_{i}$, and the resource allocation depends on the target
error required.

\subsection{Minimizing the resource at fixed error}

Suppose, instead, that we wish to minimize the complexity for a fixed
error requirement. An experimental measurement may have a known error-bar,
and one wishes to compare theory with experiment. In such cases, obtaining
the theoretical prediction with an error much lower than experimental
errors is wasteful of computational resources.

To minimize the resource $N$ at fixed total error $\epsilon$, we
minimize:
\begin{equation}
N\left(\epsilon,\lambda\right)=\prod N_{i}+\lambda_{N}\left(\epsilon\left(\bm{N}\right)-\hat{\epsilon}\right),
\end{equation}
where the Lagrange multiplier $\lambda_{N}$ is chosen so that there
is a predetermined value of $\hat{\epsilon}=\epsilon\left(\bm{N}\right)$.
Differentiating again, 
\begin{equation}
\frac{\partial N\left(\epsilon,\lambda\right)}{\partial N_{i}}=\frac{1}{N_{i}}\left(N-\lambda_{c}pn_{i}\epsilon_{i}\left(N_{i}\right)\epsilon\left(\bm{N}\right)^{1-p}\right)=0
\end{equation}

This is the same equation as before, except with $\lambda_{N}=1/\lambda_{\epsilon}$.
Since the value of the Lagrange multiplier is eliminated from the
solution, the scaling is unchanged. The minimum resource usage $N$
for a total error $\epsilon$ is obtained when:
\begin{equation}
N=\left(\epsilon/\epsilon_{0}\right)^{-1/c}.
\end{equation}

\section{Differential equation errors\label{sec:Stochastic-integration-errors}}

Standard definitions for ordinary differential equation solvers define
the order so that the error is a power of the step-size in time. Our
definitions in Section (\ref{sec:Complexity-order}) are consistent
with this. Since the required resources $N$ are inverse to the time-step,
one has $\epsilon=\epsilon_{0}N^{-n}\propto\Delta t^{n}.$ For an
ordinary differential equation, our definition of complexity order
agrees with the usual definition for a one-step algorithm with a global
error of $\epsilon$ for a time-step $\Delta t$, that is, $n=c$.
However, there are many computational problems with multiple error
sources, and an optimization is necessary to obtain optimal efficiency.

As an example, the overall error in a numerical solution of a stochastic
differential equation includes both the time-step error and the sampling
error. Most numerical analyses focus on the step-size error \citep{ruemelin1982numerical,Kloeden1992numerical,breuer2000numerical,tocino2002runge,wilkie2004numerical,jentzen2009numerical,burrage2006comment},
which is defined for an infinite ensemble. This ignores the sampling
error which occurs as well, and also contributes to the complexity
order. For stochastic differential equation (SDE) solvers , one is
solving an equation of form
\begin{equation}
dx=a(x)dt+\sum_{j=1}^{d}b_{j}\left(y\right)\circ dW_{j},
\end{equation}
where $x$ is an n-dimensional real or complex function of time, $dW_{j}$
is a vector of real Gaussian noises, and the product notation $\circ$
indicates the use of Stratonovich calculus, for definiteness.

The usual analyses assume that the desired quantity is either a probability
averaged over an infinite number of samples, called 'weak' convergence,
or else a sample with a known noise, called 'strong' convergence \citep{Kloeden1992numerical,burrage2006comment,jentzen2009numerical}.
Different orders are possible, depending on the method. Yet there
is another error source, which is the sampling error.

Similar combinations of errors due to different resources are found
for partial differential equations and stochastic partial differential
equations. In this paper, we focus on probabilities or moments, to
obtain a strategy giving the lowest total error for the total resource
used.

\subsection{Stochastic errors}

Either weak or strong convergence analysis allows the treatment of
the time-step error $\epsilon_{T}$, such that $\epsilon_{T}\propto\Delta t^{n}.$
In most practical applications, one often wishes to estimate a probabilistic
quantity or average, given only a finite set of samples, since it
is not possible to obtain an infinite ensemble.

In these cases, if the stochastic trajectory is $x(t)$ , one wishes
to evaluate a function 
\begin{align*}
\bar{g}(t)=\left\langle g\left(x(t)\right)\right\rangle _{\infty} & =\lim_{N_{S}\rightarrow\infty}\frac{1}{N_{S}}\sum_{i=1}^{N_{S}}g\left(x^{(i)}(t)\right).
\end{align*}

The total error is a combination of the time-step error and the sampling
error of a finite ensemble. These combine in quadrature, since the
result for a computational set of $N_{S}$ trajectories has the form,
for a residual sampling error $\Delta w$:
\begin{equation}
\bar{g}_{N_{S}}(t)=\bar{g}_{c}(t)+k\Delta t^{n}+\Delta w\left(t\right).
\end{equation}

Here, $\bar{g}_{N_{S}}(t)$ is the computed moment at time $t$. This
may correspond to the original stochastic variable or a function of
it. We define $\bar{g}_{c}$ as the correct or targeted mean value
in the infinite ensemble limit. As a result, we have weak convergence,
such that:
\begin{equation}
\left\langle g\left(t\right)\right\rangle _{\infty}=\bar{g}_{c}(t)+k\Delta t^{n}
\end{equation}
However, the resulting complexity is not of much practical interest.
An infinite set of samples is impossible, and even using more samples
than necessary wastes computational time and energy.

The simplest method of sampling in a finite ensemble is to perform
$N_{S}$ repeats that each involve $N_{T}$ time-steps, requiring
resources of $N_{A}$ per algorithmic time-step, with independent
random noises. The results of most interest are the averages over
the $N_{S}$ random trajectories. The total resources used are therefore
$N=N_{A}N_{T}N_{S}.$ For trajectories which are non-Gaussian, mean
values and variances are optimally calculated in two stages \citep{Kiesewetter2016xSPDE,opanchuk2018simulating}.
This is often more efficient numerically, since it allows better use
of parallel computation, but it can still be carried out in series.

In such cases, one has $N_{S}=N_{S}^{(1)}N_{S}^{(2)}$, and from the
central limit theorem \citep{billingsley2017probability}, if a moment
or sampled probability is first calculated using the mean of the sub-ensemble
$N_{S}^{(1)}$, the computed results have a Gaussian distribution
at large $N_{S}^{(1)}$. After a final average, they have an error
in the overall mean that can be estimated from the variance in the
$N_{S}^{(2)}$ ensemble, proportional to $1/N_{S}^{(2)}.$

The result of the analysis is that the error estimate for an SDE algorithm
with a time-step error order $n$ and sampling order $s$, where typically
$n\ge1$ and $s=1/2$, is:
\begin{align}
\epsilon & =\sqrt{\epsilon_{T}^{2}+\epsilon_{S}^{2}}=\sqrt{\epsilon_{0T}^{2}N_{T}^{-2n}+\epsilon_{0S}^{2}N_{S}^{-2s}}.
\end{align}

\subsection{Fixed resource strategies}

We now consider how the total error scales with increased resources,
for the two-resource case of a stochastic differential equation, with
$N=N_{A}N_{T}N_{S}$. Since both $N_{T}$ and $N_{S}$ can be varied
independently, we analyze three strategies one might follow. An optimal
strategy that is better than any of these is treated next.

Conventional error analysis typically supposes that one fixes one
resource, while varying the other one. For a stochastic differential
equation, one may fix $N_{S}$, and vary the time-step so $N_{T}$
increases. The approach is justified by assuming an infinite number
of samples, but in reality the number of samples is always finite
and the sampling error is the largest term at small step-size:
\begin{equation}
\lim_{N\rightarrow\infty}\epsilon=\epsilon_{0S}N_{S}^{-s}.
\end{equation}
This is not a convergent strategy at large $N$, because $N_{S}$
is held constant by assumption. Similarly, one could fix the step-size
so that $N_{T}$ is constant, and only vary $N_{S}$. Again, the limiting
error for infinite resource utilization is not zero, but is the step-size
error, which now becomes the largest term:
\begin{equation}
\lim_{N\rightarrow\infty}\epsilon=\epsilon_{0T}N_{T}^{-n}.
\end{equation}
This is also not convergent at large $N$ values, since $N_{T}$ is
now held constant.

\subsection{Fixed resource ratios}

Another possibility is to fix the resource use ratio $r=N_{T}/N_{S}$
as $N$ increases. In this case:
\begin{align}
N_{T} & =\sqrt{rN/N_{A}}\\
N_{S} & =\sqrt{N/(N_{A}r)}.\nonumber 
\end{align}
 As a result, the step-size error and sampling error both reduce to
zero:
\begin{equation}
\epsilon=\sqrt{\epsilon_{0T}^{2}\left(rN/N_{A}\right)^{-n}+\epsilon_{0S}^{2}(rN_{A}/N)^{s}}.
\end{equation}

Since the step-size convergence order is typically $n\ge1>s$, for
this strategy, 
\begin{equation}
\lim_{N\rightarrow\infty}\epsilon=\epsilon_{0S}\left(rN_{A}\right)^{s/2}N{}^{-s/2}.\label{eq:Fixed_ratio_result}
\end{equation}
With this approach, convergence is achieved with a complexity order
of $c=1/4$ if $s=1/2$.

In summary, if one reduces both the sampling and step-size errors
with a constant resource ratio, the step-size error becomes negligible
at large resource usage compared to the sampling error. Using a high
order technique is of little utility here. The errors are due to sampling,
not step-size error, in the limit of large resource use.

With a constant resource ratio strategy, it is advantageous to use
a method that is fast and efficient. The reason is clear from Eq (\ref{eq:Fixed_ratio_result}),
which shows that for fixed resources $N$, the error $\epsilon$ increases
with the step complexity $N_{A}$. In achieving a given target error,
the only effect of higher-order methods is to increase the resource
requirement.

\section{SDE\label{sec:Stochastic-complexity-order} complexity order}

Is there any asymptotic advantage to using a stochastic method for
an SDE with a higher step-size order? In this section we show in detail
that there is an advantage if a more sophisticated optimization is
used. This is feasible, but we show that the complexity order improvements
are less than one might hope for. Our derivation gives the same result
as the general argument, but for two resource components a direct
proof is possible without Lagrange multipliers.

\subsection{Minimizing the errors}

The optimal approach is to vary the ratio $r$, changing this with
the total resources $N$ fixed, so as to minimize the total error.
To simplify the equations, and give a more intuitive result, define
$\epsilon_{T}\left(r_{0}\right)$ as the step error and $\epsilon_{S}\left(r_{0}\right)$
as the sampling error at a given resource ratio $r_{0}$. It follows
that the total error is given by
\begin{equation}
\epsilon\left(r\right)=\sqrt{\epsilon_{T}^{2}\left(r_{0}\right)\left(r/r_{0}\right)^{-n}+\epsilon_{S}^{2}\left(r_{0}\right)(r/r_{0})^{s}}.\label{eq:Total_error_4}
\end{equation}

Assuming that $s=1/2$, one has a minimum total error when $\left[\epsilon_{S}^{2}\left(r\right)-2n\epsilon_{T}^{2}\left(r\right)\right]=0.$
The optimum ratio for given computational resources $N$ is therefore
obtained when the sampling error to step size error ratio is fixed,
in agreement with Eq (\ref{eq:resource power ratio}). This gives
a larger sampling than step-size error, with an error ratio of:
\begin{equation}
\frac{\epsilon_{S}\left(r\right)}{\epsilon_{T}\left(r\right)}=\sqrt{2n}.\label{eq:es_et}
\end{equation}

There is a simple, intuitive explanation. Since the step-size error
varies fastest with resource usage, a smaller step-error is effective
at balancing a larger sampling error, with the ratio depending on
the order. However, a step-size error much lower or higher than the
optimum is not efficient, as it wastes computational resources.

From Eq (\ref{eq:Total_error_4}), this result corresponds to having
a resource ratio of:
\begin{equation}
r=r_{0}\left[\frac{2n\epsilon_{T}^{2}\left(r_{0}\right)}{\epsilon_{S}^{2}\left(r_{0}\right)}\right]^{1/(n+1/2)}.
\end{equation}
Recalling that $\tilde{N}=N_{T}N_{S}=N/N_{A}$, this is in agreement
with the variational result of Eq (\ref{eq:variational_ratio}).

On solving for the total error estimate, one finds that there is
a power law in $N$, in agreement with the Laplace multiplier results,
i.e., $\epsilon=\epsilon_{0}N^{-c}$. Convergence is achieved in this
optimum allocation with a complexity order identical to that obtained
in Sec (\ref{sec:Complexity-order}):
\begin{align}
c & =\frac{n}{2n+1}.\label{eq:complexity_order_prediction}
\end{align}
and a leading coefficient equal to that of Eq (\ref{eq:complexity_prefactor}):
\begin{equation}
\epsilon_{0}=\sqrt{2+1/n}N_{A}^{c}\left(\frac{n^{\frac{1}{2}}\epsilon_{0T}\epsilon_{0S}^{2n}}{2^{n}}\right)^{\frac{1}{1+2n}}.
\end{equation}
 Since the weak error time-step convergence of a stochastic equation
has a typical range of $1\le n<\infty$ for the SDE case, one finds
that $1/3\le c<1/2.$ This generalizes results on resource optimization
that are also known in the financial mathematics literature \citep{duffie1995efficient}.
It applies to more sophisticated multi-level solvers as well \citep{giles2008multilevel,giles2015multilevel,haji2016multi}.

\subsection{Implications of stochastic complexity order}

With the optimal strategy, the complexity order is always higher,
and the algorithm always converges faster than with a fixed ratio
strategy. The maximum complexity order is $0.5$, and the order varies
relatively slowly with the algorithmic step-size order.

In practical applications, suppose that $n$ is known, and one obtains
$\epsilon_{T}$ and $\epsilon_{S}$ to satisfy this equation at some
$N'_{T}$ and $N'_{S}$. Then the optimum is obtained at all resource
allocations, provided the equality $\epsilon_{S}\left(r\right)=\sqrt{2n}\epsilon_{T}\left(r\right)$
can be maintained. This implies that on changing the resource by a
factor of $\lambda$, one must ensure that:
\begin{align}
N_{S} & =\alpha N'_{S};\,N_{T}=\alpha^{1/2n}N'_{T}.
\end{align}
 where the factor $\alpha$ is defined so that $\alpha=\lambda^{2c}.$
This shows that more resources should be used to minimize the sampling
error than the step-size error, providing the most effective use of
the computational time and energy.

\subsection{Kubo oscillator example \label{sec:Numerical-examples}}

As an example of an SDE with two resources, consider the Kubo oscillator
\citep{kubo1954note,Anderson1954mathematical}, which describes a
physical oscillator with a random frequency. This has a wide applicability
in physics, chemistry, biology and economics \citep{jung2002stochastic,sato1998dynamic,turelli1977random}.
It is described by the following stochastic differential equation:
\begin{equation}
dx=i\omega_{0}xdt+ix\circ dw,
\end{equation}
using Stratonovich \citep{stratonovich1960theory} calculus, where
$dw$ is a real Gaussian noise such that $\langle dw^{2}\rangle=dt$.
In the Ito calculus \citep{ito1996diffusion,Gardiner2009Stochastic},
the equivalent stochastic equation is
\begin{equation}
dx=(i\omega_{0}-0.5)xdt+ixdw,\label{eq:Kubo_Ito}
\end{equation}
The expectation value for the moment $x^{m}$ has an analytical solution
\citep{Drummond1991Computer}:
\begin{align}
\langle x^{m}(t)\rangle & =\langle x^{m}\left(0\right)\rangle e^{-t/2\left(m^{2}-2im\omega_{0}\right)}\,.
\end{align}

We will treat methods with different time-step orders. The first is
a midpoint method (MP) with first-order weak convergence. This semi-implicit
method is useful for stiff differential problems and stochastic partial
differential equations \citep{Drummond1991Computer,Werner1997Robust},
due to its stability. The second is a fourth-order Runge-Kutta method,
(RK4) which gives second-order weak stochastic convergence in this
case, although not for all cases \citep{wilkie2004numerical,burrage2006comment,press2007numerical}.
Both of these are used for Stratonovich calculus. The third is a second
order weak stochastic Runge-Kutta method (KPW2), designed for Ito
stochastic differential equations \citep{Kloeden1992numerical}. 

\begin{figure}[h]
\begin{centering}
\par\end{centering}
\begin{centering}
\includegraphics[width=0.9\columnwidth]{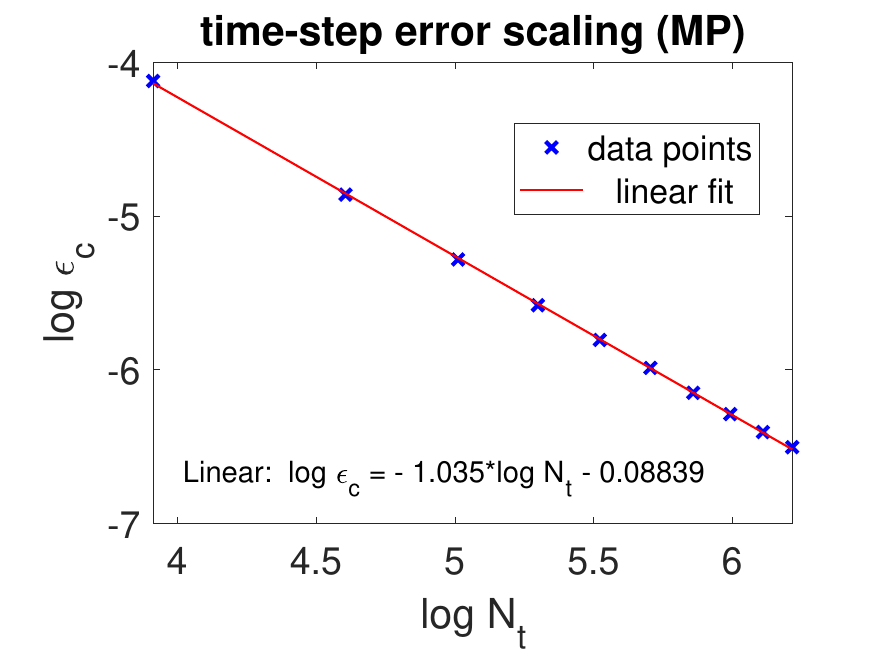}
\par\end{centering}
\begin{centering}
\includegraphics[width=0.9\columnwidth]{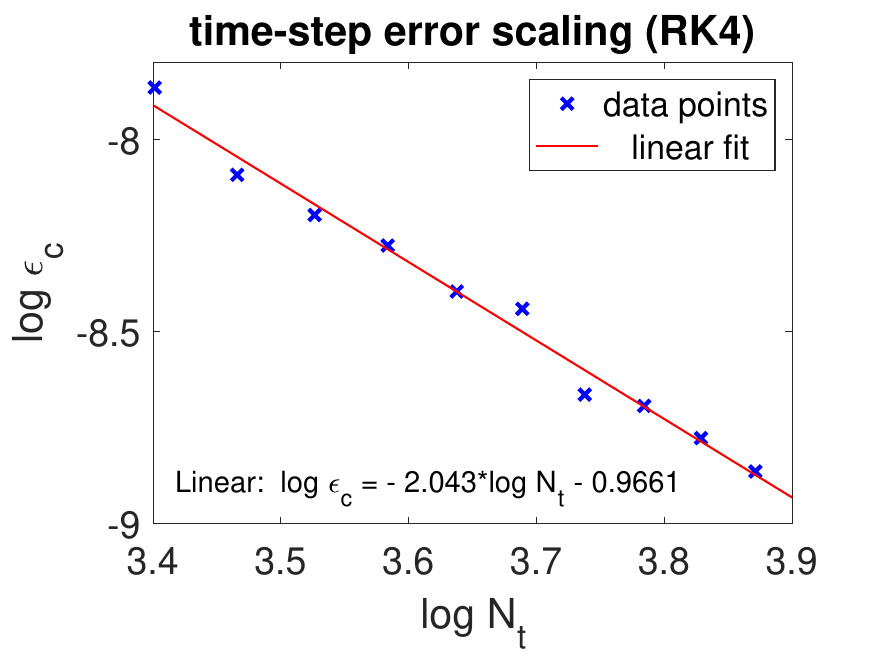}
\par\end{centering}
\begin{centering}
\includegraphics[width=0.9\columnwidth]{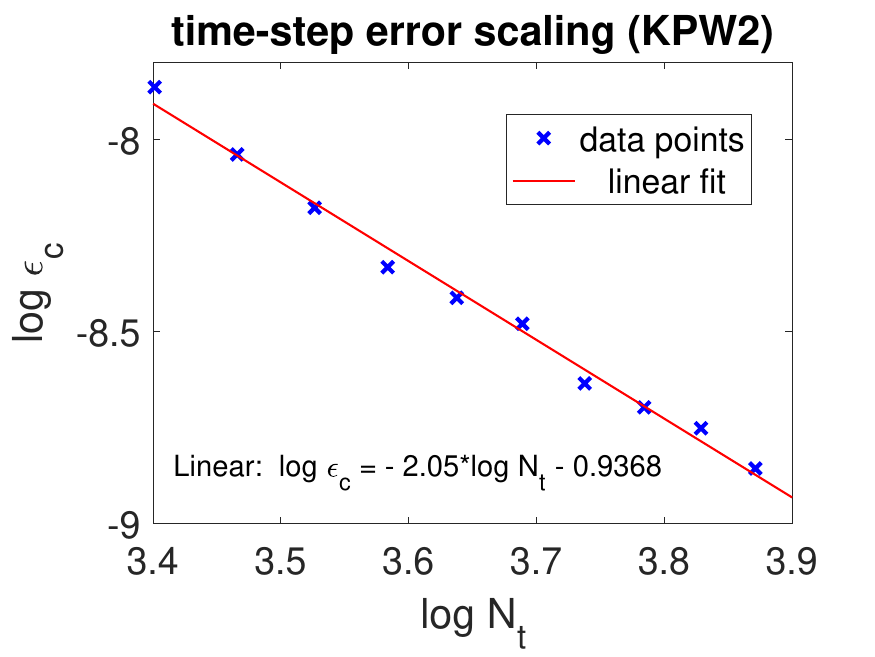}
\par\end{centering}
\caption{The error against number of time-steps plots for an SDE using the
midpoint (MP) method (top) , Runge-Kutta (RK4) method (middle) and
weak second order (KPW2) method (bottom). The time-step error scalings
estimated from these plots are tabulated in Table \ref{tab:sampling_error_scaling_results}.\label{fig:timestep_error_scaling_plots}}
\end{figure}

\begin{figure}[h]
\begin{centering}
\par\end{centering}
\begin{centering}
\includegraphics[width=0.9\columnwidth]{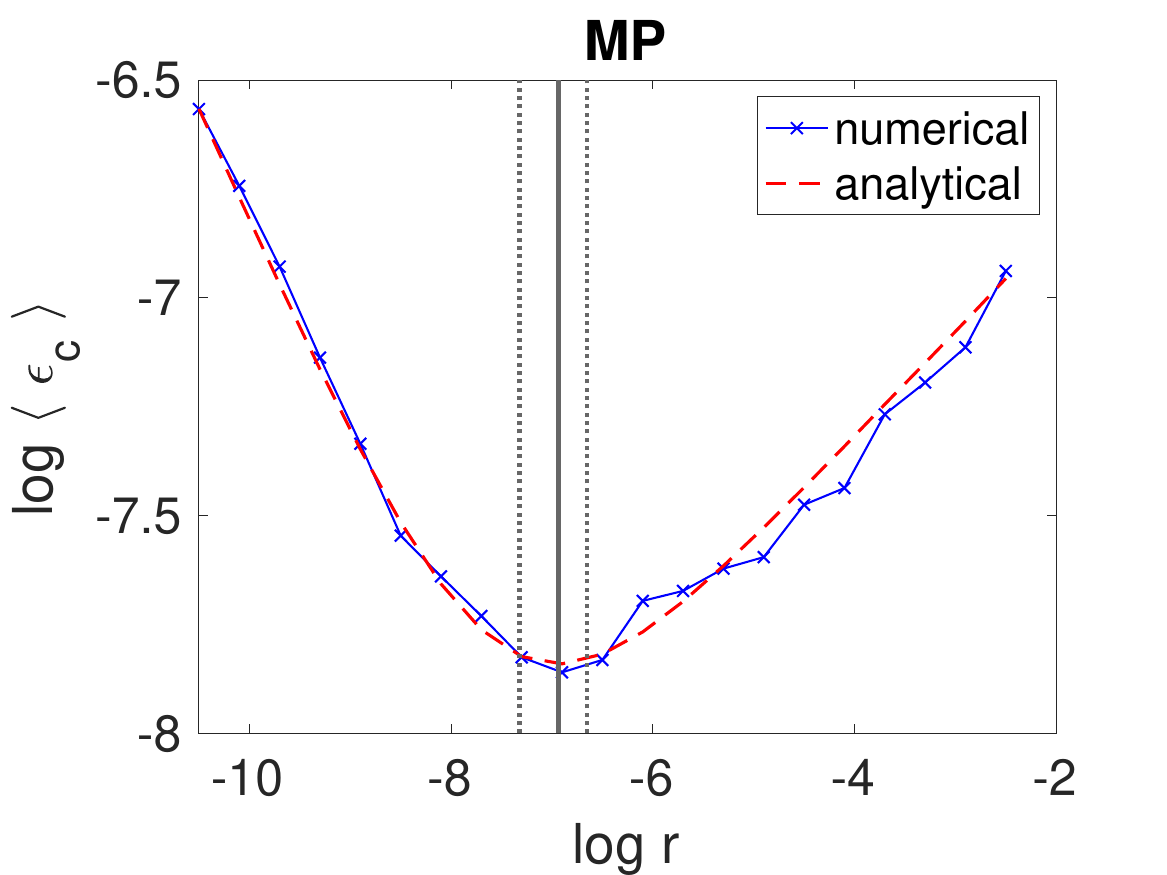}
\par\end{centering}
\begin{centering}
\includegraphics[width=0.9\columnwidth]{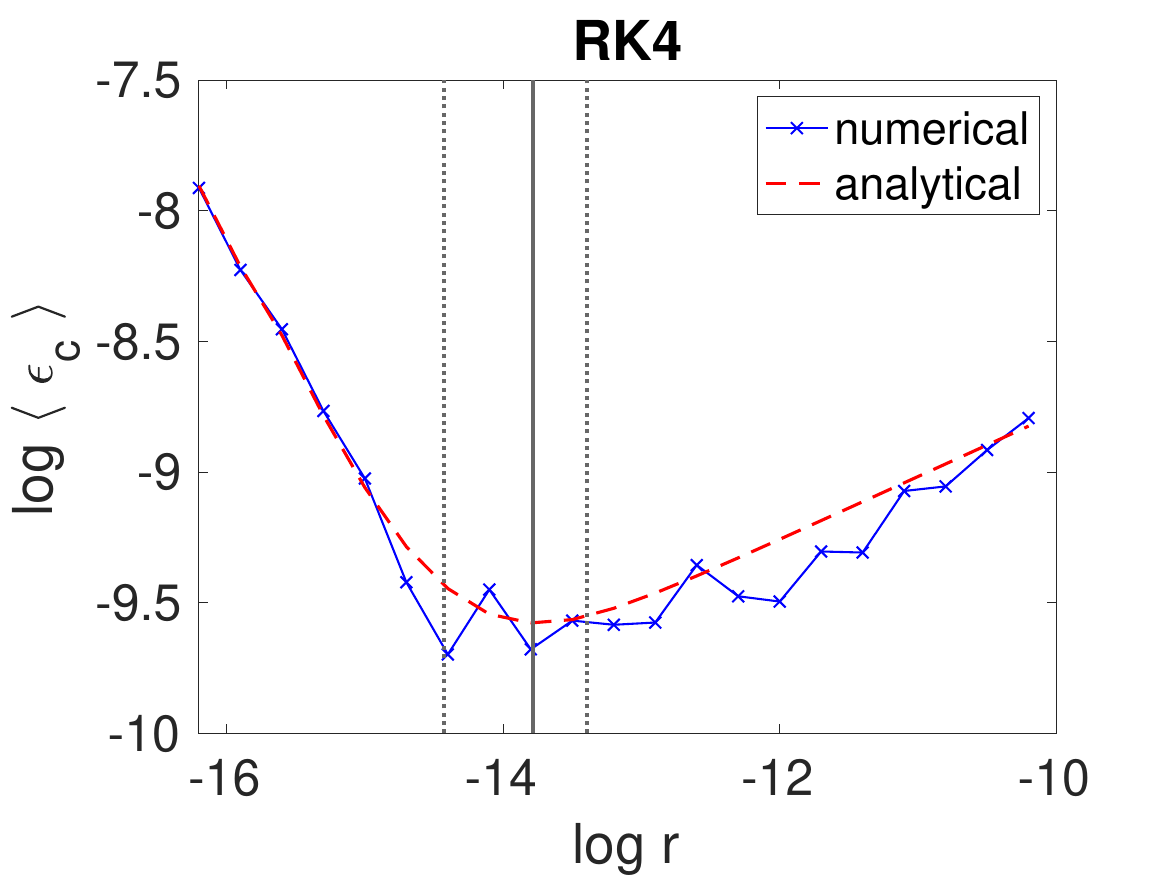}
\par\end{centering}
\begin{centering}
\includegraphics[width=0.9\columnwidth]{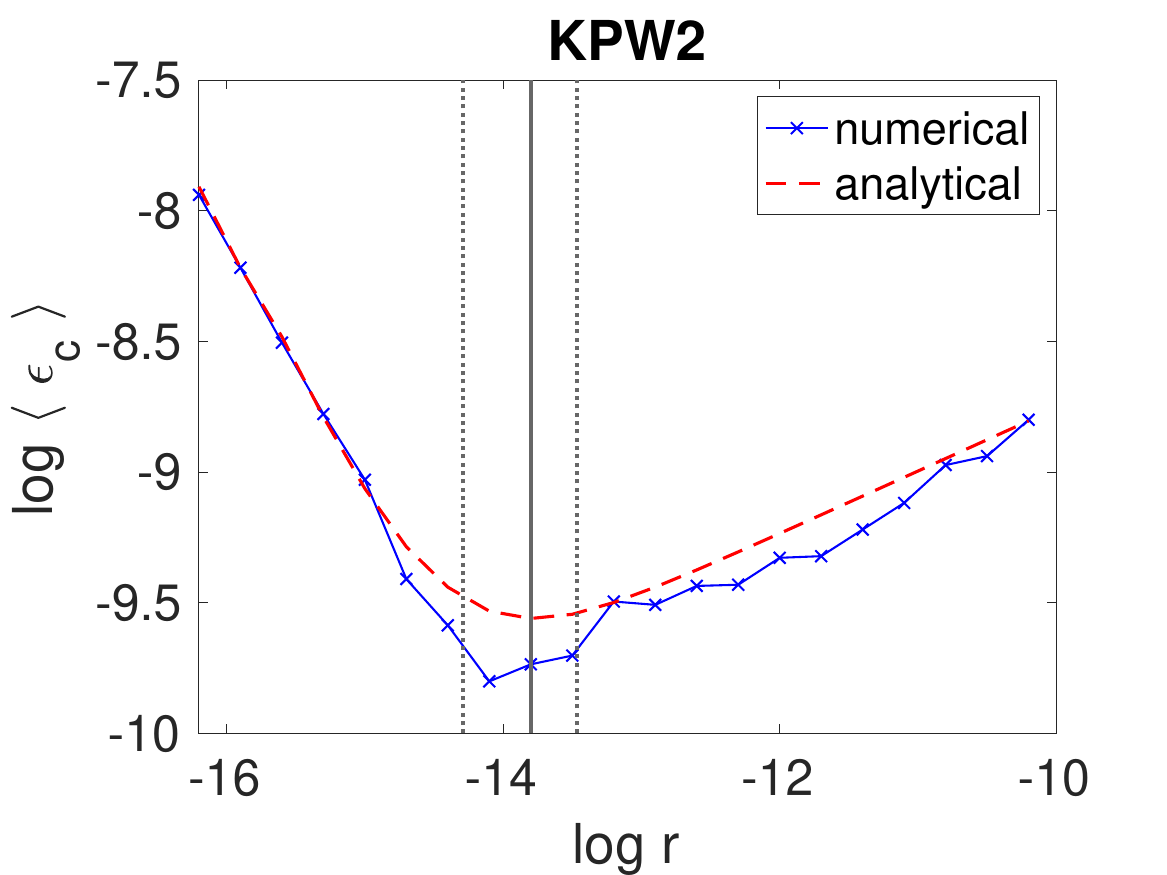}
\par\end{centering}
\caption{Here $\bar{\epsilon}_{c}$ is the average comparison error , and $r=N_{t}/N_{s}$
is the ratio between the time points and samples for a total resource
of $N=10^{10}$. The numerical results are the blue solid curves;
the red dashed curves are the theoretical total errors of Eq. (\ref{eq:total error}).
The optimal ratios of Eq. (\ref{eq:variational_ratio}) are the solid
grey lines, the error-bars from Eq. (\ref{eq:uncertainty}) are the
dashed lines. The graphs are for the MP (top), RK4 (middle) and KPW2
(bottom) methods. \label{fig:comparison_error_scaling_plots}}
\end{figure}

The numerical example is the Kubo oscillator with $\omega_{0}=0$
, where we compute $\langle x(t)\rangle$, with an initial state $x(0)=1$.
The error scalings for an algorithm have to be determined to obtain
the complexity order. We evaluate the sampling order and step-size
order in Table \ref{tab:sampling_error_scaling_results} and Fig.
(\ref{fig:timestep_error_scaling_plots}), then compute the observed
complexity order in Table \ref{tab:complexity_order_results} and
Fig. (\ref{fig:comparison_error_scaling_plots}). Finally, we compare
our results with the expected asymptotic complexity orders for all
three methods. For simplicity, we assume $N_{A}=1$ in all cases,
although there are small variations in the resource per algorithm.

We used two codes for cross-checking. The first was a public domain
code for solving stochastic (and partial\} differential equations
named xSPDE4 \citep{Kiesewetter2016xSPDE,kiesewetter2023xspde3,Drummond2024xSPDE4}.
This computes and estimates the sampling error described above, as
well as other errors. All results were checked against a second independently
written code, which gave excellent agreement.

\begin{table*}
\begin{centering}
\begin{tabular}{|c|c|c|c|c|}
\hline 
 & time-step exponent, $n$ & $log_{e}(\epsilon_{0T})$ & sampling exponent $s$ & $log_{e}(\epsilon_{0S})$\tabularnewline
\hline 
\hline 
MP & $1.035\pm0.004$ & $-0.09\pm0.02$ & $0.484\pm0.011$ & $-0.78\pm0.17$\tabularnewline
\hline 
RK4 & $2.04\pm0.09$ & $-0.97\pm0.31$ & $0.483\pm0.011$ & $-0.80\pm0.18$\tabularnewline
\hline 
KPW2 & $2.05\pm0.07$ & $-0.94\pm0.25$ & $0.480\pm0.011$ & $-0.83\pm0.17$\tabularnewline
\hline 
\end{tabular}~
\par\end{centering}
\caption{The time-step error scalings and $log_{e}\left(\epsilon_{0T}\right)$
values for the MP, RK4, and KPW2 methods for an SDE, followed by the
sampling error scalings and $log_{e}\left(\epsilon_{0S}\right)$ values.
The error bars are the standard deviation in the mean (see Appendix).
\label{tab:sampling_error_scaling_results}}
\end{table*}

\subsection{Time-step error scaling}

We estimate the time-step error scaling, $n$, from the comparison
error $\epsilon_{c}$ assuming that $\epsilon_{c}=\epsilon_{0T}N_{t}^{-n}$,
for$N_{t}$ time steps. The time range for the simulation is $5$.
Similar to the sampling error scaling estimation, we graphed $log_{e}\,\epsilon_{c}$
against $log_{e}\,N_{t}$, where $\epsilon_{c}$ is the comparison
error, and estimate the time-step error scaling from the gradients
of these plots. First, the mean of means $\bar{x}(t_{k})$ at time
$t_{k}$ of the Kubo oscillator is computed (see Eq. (\ref{eq:mean_of_means})),
for all time points $k=1,...,N_{t}$.

The comparison error is defined as the scaled root-mean-square (RMS)
difference between the computed and exact values:
\begin{align}
\epsilon_{c} & \equiv\frac{1}{\bar{x}_{max}}\sqrt{\frac{1}{N_{t}}\sum_{k=1}^{N_{t}}\left(\bar{x}(t_{k})-x_{exact}(t_{k})\right)^{2}}\,.\label{eq:comparison_error}
\end{align}

These plots are presented in Fig. \ref{fig:timestep_error_scaling_plots}.
For the midpoint (MP) method, we chose a range of $50$ to $500$
time points, which implies a largest time-step of less than $0.1$,
and a fixed $2\times10^{9}$ number of samples. The time-step error
scaling exponent was $1.035\pm.004$. While slightly different to
the expected $n=1$, there are approximations in the fitting methods
(see Appendix), which may cause this.

For both the Runge-Kutta (RK4) and weak second order (KPW2) methods,
we used $30$ to $48$ time points, giving a largest time-step of
less than $0.17$, and a fixed number of $2\times10^{9}$ samples.
We chose more time points for the MP method compared to the other
two methods, given their smaller errors relative to the MP method.
The results are tabulated in Table \ref{tab:sampling_error_scaling_results}.
The plots are presented in Fig. \ref{fig:timestep_error_scaling_plots}.

\subsection{Sampling error scaling}

Next we estimate the sampling error scaling, $s$, from the sampling
error $\epsilon_{s}=\epsilon_{0}N_{s}^{-s}$. We fixed the number
of time steps to be $N_{t}=5000$ with a simulation time range of
$5$, to give a time-step of less than $0.001$. The sample sizes
$N_{s}^{(2)}$ were chosen to range from $2000$ to $4000$, while
$N_{s}^{(1)}=2000$ is fixed. This gives a total number of samples
$N_{S}=N_{S}^{(1)}N_{S}^{(2)}$ ranging from $4\times10^{6}$ to $8\times10^{6}$.

The sampling error is computed based on the two-stage description
in Sec. \ref{sec:Stochastic-integration-errors}, with the number
of samples $N_{S}=N_{S}^{(1)}N_{S}^{(2)}$. The overall mean $\bar{x}$
and sub-ensemble means $\bar{x}_{i}\left(t_{k}\right)$ are given
respectively by
\begin{align}
\bar{x}_{i}\left(t_{k}\right) & =\frac{1}{N_{s}^{(1)}}\sum_{j=1}^{N_{s}^{(1)}}x_{ij}\left(t_{k}\right)\label{eq:mean_of_means}\\
\bar{x}\left(t_{k}\right) & =\frac{1}{N_{s}^{(2)}}\sum_{i=1}^{N_{s}^{(2)}}\bar{x}_{i}\left(t_{k}\right)\nonumber 
\end{align}
at time $t_{k}$, where $k=1,...,N_{t}$. The variance of the overall
mean at time $t_{k}$ is 
\begin{align}
\Delta^{2}x\left(t_{k}\right) & =\frac{1}{N_{s}^{(2)}}\sum_{i=1}^{N_{s}^{(2)}}\left(\bar{x}_{i}^{2}(t_{k})-\bar{x}(t_{k})^{2}\right)\,,
\end{align}
while the standard error $\sigma\left(t_{k}\right)$ at time $t_{k}$
and root-mean-square (RMS) error $\epsilon_{s}$ over all times are
given respectively by
\begin{align}
\sigma\left(t_{k}\right) & =\sqrt{\frac{\Delta^{2}x\left(t_{k}\right)}{N_{s}^{(2)}}}\nonumber \\
\epsilon_{s} & =\frac{1}{\bar{x}_{max}}\sqrt{\frac{1}{N_{t}}\sum_{k=1}^{N_{t}}\sigma^{2}\left(t_{k}\right)}\,.
\end{align}

Here $\bar{x}_{max}$ is the maximum value of $\bar{x}(t_{k})$ over
all time points, and is used to give a dimensionless relative error.
We evaluated $log_{e}\,\epsilon_{s}$ as a function of $log_{e}\,N_{s}$,
where $\epsilon_{s}$ is the sampling error. Using least squares fitting
\citep{acton1966analysis,bevington2003data}, we estimate the sampling
error scaling from the gradients of these plots, while the y-intercepts
were also recorded for the complexity order estimations. The standard
errors of these quantities are also computed \citep{acton1966analysis,bevington2003data}
(refer to the Appendix for the exact mathematical expressions). These
results are tabulated in Table \ref{tab:sampling_error_scaling_results}.
The slight reduction in order of $0.02\pm0.01$ compared to the expected
value of $0.5$ is due to residual effects of non-zero time-step errors.
In all cases, there was an excellent fit.

\subsection{Numerical complexity optimization}

In the complexity order estimations in this subsection, the optimal
ratio between the number of samples and time points is used. This
ratio is determined by the expression in Eq. (\ref{eq:variational_ratio})
and should yield a minimum comparison error. We first verify this
in two different ways; one is analytic and the other numerical. A
check on the relation $\epsilon_{s}/\epsilon_{t}=\sqrt{n/s}$ (Eq.
(\ref{eq:es_et})) that holds when the optimal ratio is used serves
as an analytic validation. For a given total resource, the optimal
ratio dictates the optimal sample size $N_{s,opt}$ and time points,
$N_{t,opt}$ which determine the sampling error $\epsilon_{s}=\epsilon_{0s}N_{s,opt}^{-s}$
and time-step error $\epsilon_{t}=\epsilon_{0t}N_{t,opt}^{-n}$, respectively,
allowing $\epsilon_{s}/\epsilon_{t}$ to be calculated.

A numerical check was carried out by taking a range of ratios $r=N_{t}/N_{s}$
and computing the corresponding comparison errors, which should be
larger than the error from the optimal case, as shown in Fig (\ref{fig:comparison_error_scaling_plots}).
Each point corresponds to the overall RMS error in a Kubo oscillator
solution for a maximum time of $t=5.$ For each ratio, we repeated
the simulation $20$ times to generate $20$ comparison errors, $\epsilon_{c}$
for the RK4 and KP methods, and 40 times for the MP method. These
errors were then averaged to reduce the noise in the results. The
natural logarithm of this averaged comparison error, $log_{e}\,\langle\epsilon_{c}\rangle$
is plotted against the natural logarithm of ratio between the time
points and samples $log\,r$ in Fig. \ref{fig:comparison_error_scaling_plots}.

The numerical results are summarized as follows:
\begin{enumerate}
\item The MP method (top). Here, $21$ different ratios are taken with $r$
ranges from $2.8\times10^{-5}$ to $8.2\times10^{-2}$. The optimal
ratio predicted by Eq. (\ref{eq:variational_ratio}) has a value of
$(9.7\pm3.1)\times10^{-4}$, which corresponds to $3.2\times10^{6}$
samples and 3120 time points
\item The RK4 method (middle). Here, $21$ different ratios are taken with
$r$ ranges from $9.2\times10^{-8}$ to $3.7\times10^{-5}$. The predicted
optimal ratio is $(1.03\pm0.49)\times10^{-6}$ , which corresponds
to $9.8\times10^{7}$ samples and $102$ time points
\item The KPW2 method (bottom). Here, $21$ different ratios are taken with
$r$ ranges from $9.2\times10^{-8}$ to $3.7\times10^{-5}$. The predicted
optimal ratio is $(1.02\pm0.40)\times10^{-6}$ , which corresponds
to $9.9\times10^{7}$ samples and $101$ time points
\end{enumerate}
We verify the optimum ratio by taking a range of ratio values and
obtaining the corresponding errors from simulations. The minimum might
not be at the expected ratio, because there are statistical uncertainties.
However, we expect the optimal ratio to be close to the analytically
predicted ratio. In Fig (\ref{fig:comparison_error_scaling_plots}),
we see that the minimum errors are not at the expected ratios, because
there are statistical uncertainties. However, the optimal ratios are
close to the analytically predicted ratios.

Here, we determine the approximate uncertainty associated with the
optimal ratio given in Eq. (\ref{eq:variational_ratio})
\begin{align}
r_{opt} & =\left[\left(\frac{\epsilon_{0T}^{2}n}{\epsilon_{0}^{2}s}\right)\left(\frac{1}{N}\right)^{(n-s)}\right]^{1/(n+s)}\,,
\end{align}
where $n$, $s$ are the time-step error and sampling error scalings
respectively, and $N$ is the total resource, with $N_{A}=1$ for
simplicity. The uncertainty in $r_{opt}$ estimated from the uncertainties
associated with $n$, $s$, $\epsilon_{0T}$, and $\epsilon_{0}$,
using the error propagation method, is given by
\begin{align}
\sigma_{r_{opt}}^{2} & =\left(\frac{\partial r_{opt}}{\partial n}\right)^{2}\sigma_{n}^{2}+\left(\frac{\partial r_{opt}}{\partial s}\right)^{2}\sigma_{s}^{2}+\nonumber \\
 & +\left(\frac{\partial r_{opt}}{\partial\epsilon_{0T}}\right)^{2}\sigma_{\epsilon_{0T}}^{2}+\left(\frac{\partial r_{opt}}{\partial\epsilon_{0}}\right)^{2}\sigma_{\epsilon_{0}}^{2}.\label{eq:uncertainty}
\end{align}

In all cases, the numerical optimal ratio agrees with our analytic
prediction within statistical uncertainties. Hence, we use the analytic
optimum for numerical estimates of the complexity order, in the next
subsection.

\subsection{Overall complexity order}

\begin{figure}[h]
\begin{centering}
\includegraphics[width=0.9\columnwidth]{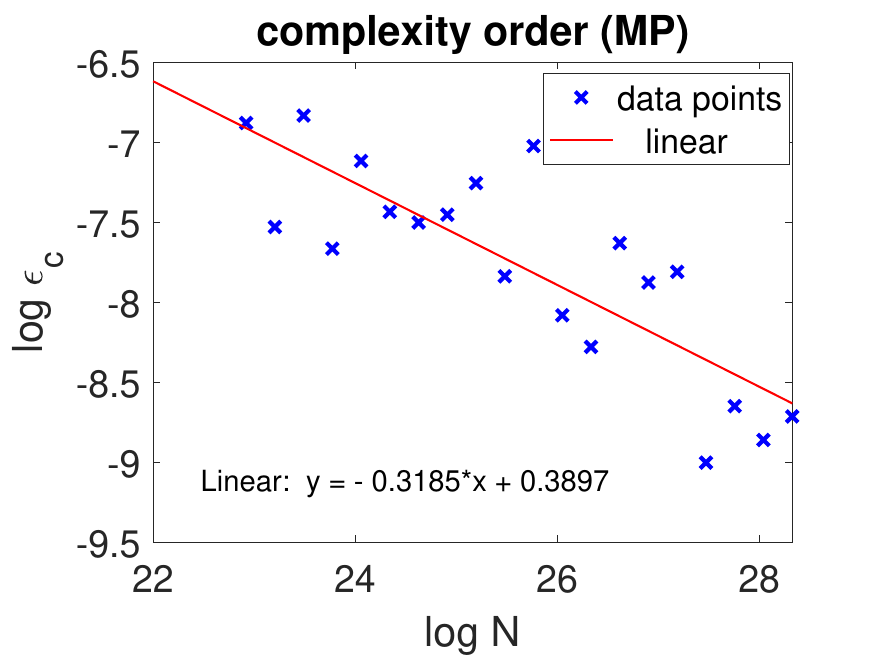}
\par\end{centering}
\begin{centering}
\includegraphics[width=0.9\columnwidth]{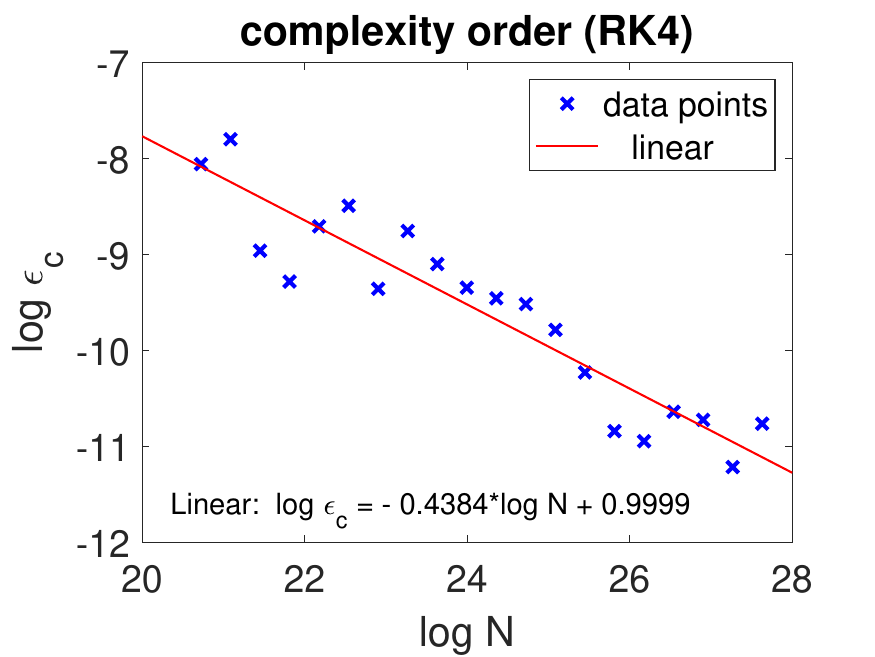}~
\par\end{centering}
\begin{centering}
\includegraphics[width=0.9\columnwidth]{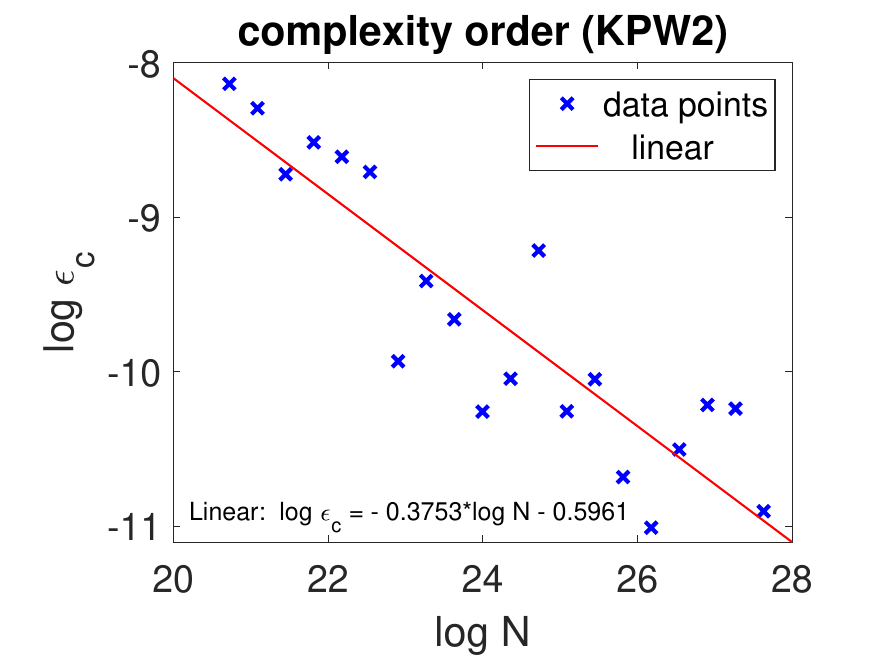}
\par\end{centering}
\caption{The error versus total resource $N$ plots for the Midpoint (MP) (top),
Runge-Kutta (RK4) (middle) and weak second order (KPW2) (bottom) method
for the Kubo SDE. \label{fig:complexity_order_plots}}
\end{figure}

With the results for sampling error and time-step error scalings (tabulated
in Table \ref{tab:sampling_error_scaling_results}), the corresponding
complexity order can be determined. This is done by using the optimum
ratio determined analytically, plotting $log_{e}\,\epsilon_{c}$ against
$log_{e}\,N$ (see Fig. \ref{fig:complexity_order_plots}), and finding
the gradient using least squares fitting. Here, $N$ is the total
resource.

For a given total resource $N$, the number of samples and number
of time points employed are determined by the ratio $r$ in Eq. (\ref{eq:variational_ratio}).
We chose a set of $20$ total resource values ranging from $9\times10^{9}$
to $2\times10^{12}$ for the midpoint (MP) method, and a set of $20$
total resource values ranging from $10^{9}$ to $10^{12}$ for both
the Runge-Kutta (RK4) and second order weak (KPW2) methods. 
\begin{table}
\begin{centering}
\begin{tabular}{|c|c|c|}
\hline 
 & complexity order, $c$ & predicted complexity order\tabularnewline
\hline 
\hline 
MP & $0.32\pm0.05$ & $0.3333$\tabularnewline
\hline 
RK4 & $0.44\pm0.04$ & $0.4000$\tabularnewline
\hline 
KPW2 & $0.38\pm0.04$ & $0.4000$\tabularnewline
\hline 
\end{tabular}
\par\end{centering}
\caption{The numerically calculated complexity order for the MP, RK4, and KPW2
methods. For the analytically predicted complexity order based on
Eq. (\ref{eq:complexity_order_prediction}), we take the asymptotic
time-step error scaling exponent for the MP method to be $1$, while
the scaling exponent for both the RK4 and KPW2 methods is $2$. \label{tab:complexity_order_results}}
\end{table}

The complexity order for the midpoint (MP) was $0.32\pm0.05$., for
the Runge-Kutta (RK4) method was $0.44\pm0.04$, while that for the
second order weak method (KPW2) was $0.38\pm0.04$, as presented in
Table \ref{tab:complexity_order_results}.

For a sampling error scaling $s$ of $0.5$, the complexity order
is obtained from the time-step error scaling using Eq. (\ref{eq:complexity_order_prediction}).
We take the time-step error scaling for the MP method to be $1$,
while the scaling for both the RK4 and KPW2 methods to be $2$. The
computed complexity orders agree with the predicted complexity orders
within the stated error bars for all methods, using two distinct computer
codes and multiple independent datasets.

\section{Partial and stochastic partial differential equations}

Other examples include partial (PDE) and stochastic partial differential
equations (SPDE), which also use multiple resources. There are errors
due to truncation in time and in each space dimension, as well as
sampling errors in the case of an SPDE. We consider resources $N_{j}$
for $j=1,\ldots d$, due to creating a lattice in a $d$-dimensional
space-time, assuming that there is no dimensional reduction due to
symmetries. In such a case, the errors depend on the algorithm. In
some cases, the errors scale independently in space and time.

For example, one may choose a method of lines in a $1+1$ space-time,
using central differencing in space combined with a Runge-Kutta method
in time \citep{holmes2007introduction}. In this case, the error is
usually estimated as a maximum, where:
\begin{align}
\epsilon & <\epsilon_{T}+\epsilon_{X}=\epsilon_{0T}N_{T}^{-n_{t}}+\epsilon_{0X}N_{X}^{-n_{x}}.
\end{align}

Hence, one has a two-resource problem that is similar to the SDE error
analysis. In typical central difference discretization approaches,
one has $n_{x}=2$ for the spatial discretization error. More generally,
there are $D$ independent resources in a $D$-dimensional space-time,
unless there is a symmetry that reduces the effective dimension. For
a stochastic partial differential equations (PSDE) there are $D+1$
resources, since there are $N_{S}$ samples needed for averaging.
The resource requirement also scales as a product of the grid size
in each dimension, giving $N=N_{A}N_{T}N_{X}N_{Y}...N_{S}$ resources
to allocate in total.

In such cases the optimal complexity order is given by the general
result of Eq (\ref{eq:optimal complexity order}), provided the errors
are additive. With more complex algorithms, there can be interactions
between the resources, as well as non-polynomial convergence properties
in the case of spectral methods \citep{boyd2001chebyshev} and sparse
grid methods \citep{Bungartz2004Sparse}. This requires a different
optimization analysis.

\subsection{Stochastic heat equation example}

Here we find the complexity order of the interaction picture midpoint
(MP) method \citep{Werner1997Robust} for solving a stochastic partial
differential equation (SPDE) that describes the stochastic diffusion
of a field $a(t,x)$ on a line. Related equations exist in many fields,
including quantum optics \citep{Carter1987Squeezing,Drummond1993Simulation},
quantum noise in atom optics \citep{Steel1998Dynamical,drummond2004canonical,Isella2005Nonadiabatic,pietraszewicz2017continuum,opanchuk2018simulating,ng2019nonlocal},
heat flow \citep{bertini1995stochastic}, fluid dynamics \citep{bell2007algorithm},
noise-driven spin systems \citep{Gao2020Nonlocal} and ecosystem and
epidemiology studies \citep{Houchmandzadeh2017Fisher}.

The SPDE treated here is given by
\begin{align*}
\frac{\partial}{\partial t}a\left(t,x\right) & =\frac{1}{2}\frac{\partial^{2}}{\partial x^{2}}a\left(t,x\right)+\eta\left(t,x\right)\,,
\end{align*}
where the noise $\eta(t,x)=\left(w_{x}+iw_{y}\right)/\sqrt{2}$ are
delta correlated in space and time, with the noise correlation $\langle w_{i}(t,x)w_{j}^{*}(t',x')\rangle=\delta(t-t')\delta(x-x')$.
The boundaries are assumed periodic in space. This is a linear equation
which is exactly soluble using Fourier transform methods. 

As a numerical test of an interaction picture SPDE algorithm \citep{Werner1997Robust},
we compute the observable $\intop\langle|a(t,x)|^{2}\rangle\,dx$,
which has an analytical solution of
\begin{align*}
\intop_{-X/2}^{X/2}\langle|a(t,x)|^{2}\rangle dx & =X\sqrt{\frac{t}{\pi}}\,,
\end{align*}
where $X$ is the spatial range. This is a three resource example,
since errors are caused by the finite time-step, finite space-step
and the finite number of independent stochastic realizations. The
algorithm uses an interaction picture with discrete Fourier transforms
solving the Laplacian part, which has an exact solution in Fourier
space. The noise term is added at the midpoint. The spatial and temporal
resources are largely independent. This allows the application of
the resource model employed here. 

By comparison, a finite difference method typically leads to a strong
coupling between the finite step-size errors in space and time, owing
to instabilities in these algorithms \citep{crank1947practical}.

\subsection{Sampling error scaling}

Just as in the previous section, we evaluated $log_{e}\,\epsilon_{s}$
as a function of $log_{e}\,N_{s}$, where $\epsilon_{s}$ is the sampling
error and $N_{s}$ is the number of samples. The time range for the
simulation is chosen to be $1$, while the spatial range is chosen
to be between $-2.5$ and $2.5$. For the sampling error scaling estimation,
we used $1001$ time points and $2000$ spatial points, which have
a time step-size $\Delta t=10^{-3}$ and a spatial stepwise $\Delta d=2.5\times10^{-3}$,
respectively. A set of computed data with values ranging from $1\times10^{3}$
samples to $2\times10^{4}$ samples is picked.

The results of the corresponding sampling error and sampling error
constant estimations are tabulated in Table \ref{tab:SPDE_error_scalings_estimations}.
\begin{table}
\begin{centering}
\begin{tabular}{|c|c|c|}
\hline 
Resource & Order & $log_{e}(\epsilon_{0})$\tabularnewline
\hline 
\hline 
Sampling & $0.56\pm0.05$ & $-0.76\pm0.47$\tabularnewline
\hline 
Time-step & $0.51\pm0.01$ & $-1.20\pm0.03$\tabularnewline
\hline 
Space-step & $1.00\pm0.06$ & $-0.96\pm0.24$\tabularnewline
\hline 
Total & $0.23\pm0.04$ & \tabularnewline
\hline 
\end{tabular}
\par\end{centering}
\caption{The sampling error scaling and $log_{e}\left(\epsilon_{0S}\right)$
value, time-step error scaling and $log_{e}\left(\epsilon_{0T}\right)$
value, and the spatial-step error scaling and $log_{e}\left(\epsilon_{0D}\right)$
value for the MP interaction picture method. The error bars are the
standard deviation in the mean (see Appendix). \label{tab:SPDE_error_scalings_estimations}}
\end{table}

\begin{figure}[h]
\begin{centering}
\includegraphics[width=0.9\columnwidth]{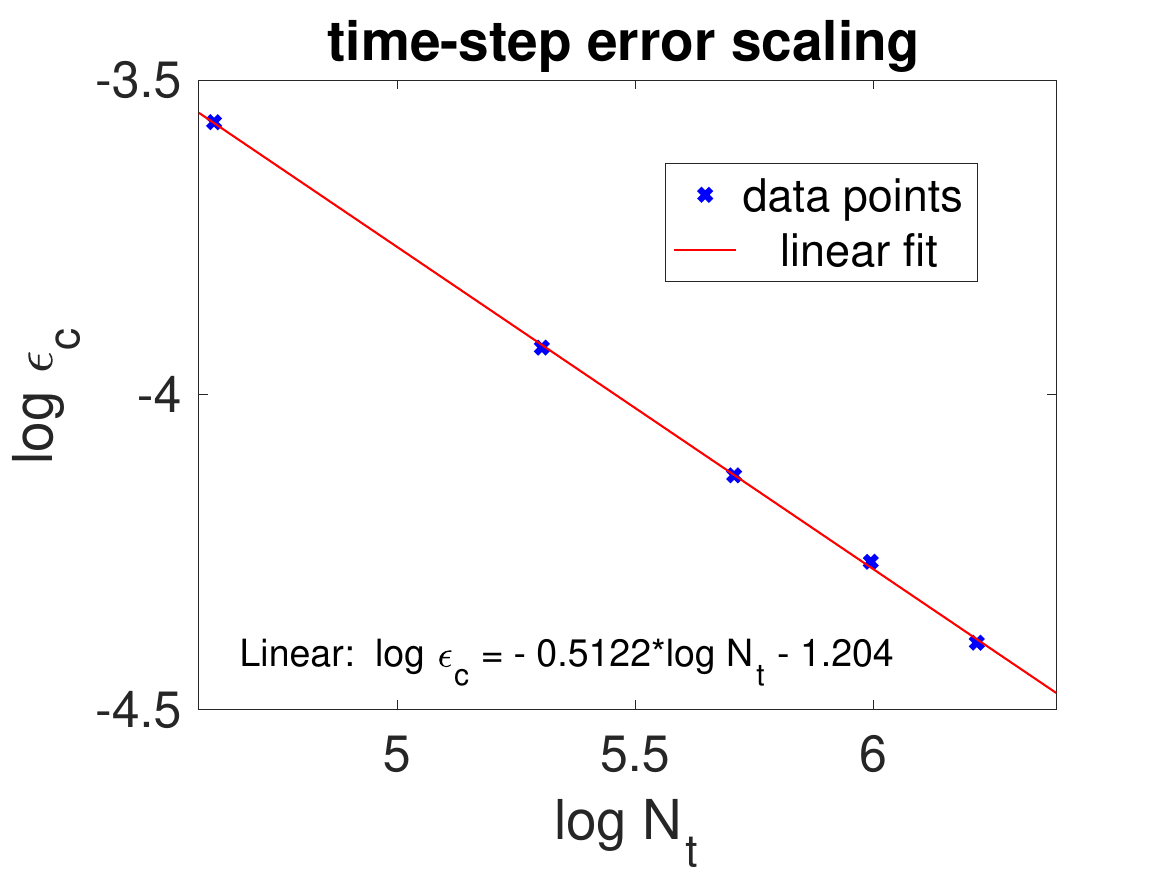}
\par\end{centering}
\begin{centering}
\includegraphics[width=0.9\columnwidth]{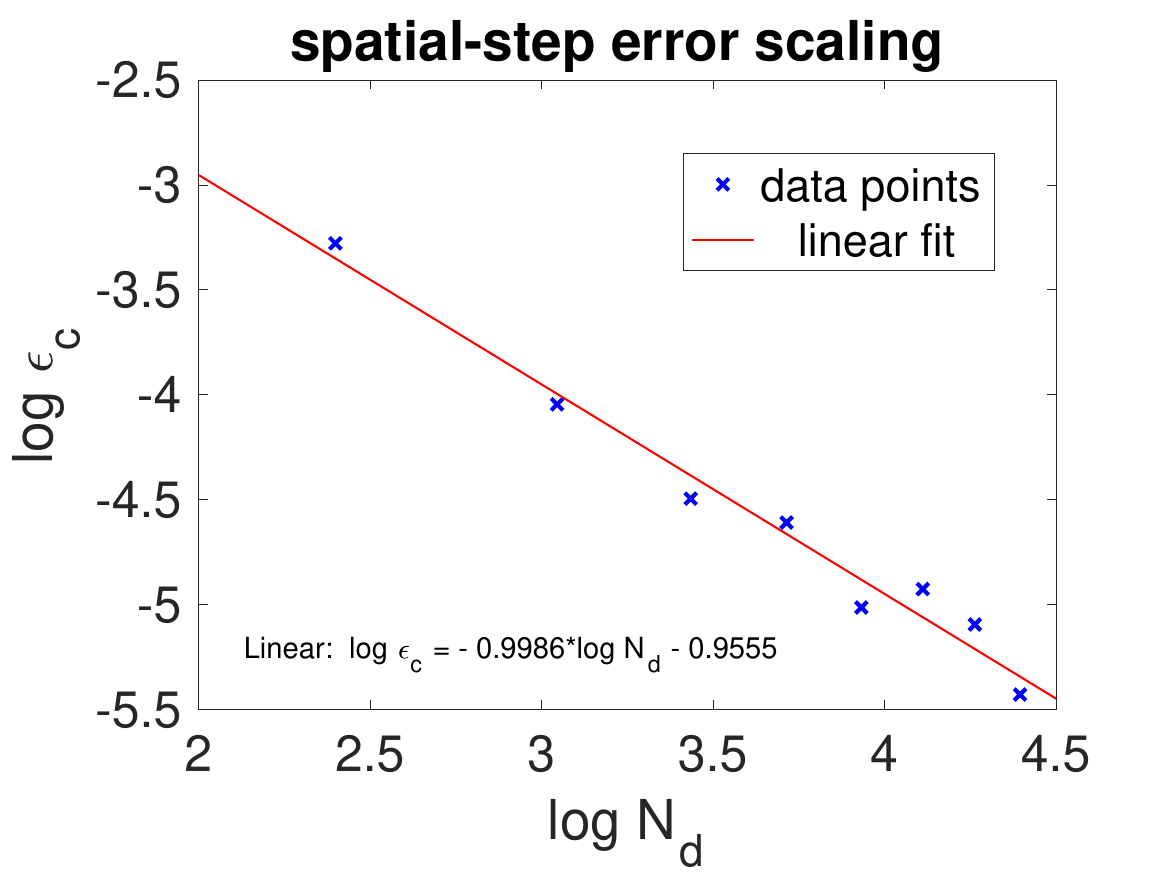}
\par\end{centering}
\caption{Time-step error (top) and space-step error (bottom), for the the midpoint
(MP) method with an SPDE. The error scalings from the plot are tabulated
in Table \ref{tab:SPDE_error_scalings_estimations}. \label{fig:step_error_planar_noise}}
\end{figure}

\subsection{Time and space-step error scaling}

To estimate the time-step error scaling, we compute the comparison
errors $\epsilon_{c}$ (defined as the root-mean-square (RMS) difference
between the computed and exact values, as in Eq. (\ref{eq:comparison_error}))
for a range of $101$ to $501$ time points $N_{t}$, while fixing
the number of samples to be $2\times10^{4}$ and $2000$ spatial points.
The result is presented in Fig. \ref{fig:step_error_planar_noise}.
The time-step error scaling exponent was $0.51\pm0.01$.

Next, we estimate the spatial-step error scaling by computing the
comparison errors $\epsilon_{c}$ for a range of spatial points $N_{d}$,
while fixing the number of samples to be $2\times10^{4}$ and $1001$
time points. The result is also presented in Fig. \ref{fig:step_error_planar_noise}.
The spatial-step error scaling exponent was $\text{1.00\ensuremath{\pm0.06}}$.

\subsection{SPDE complexity order estimation}

\begin{figure}[h]
\begin{centering}
\includegraphics[width=0.9\columnwidth]{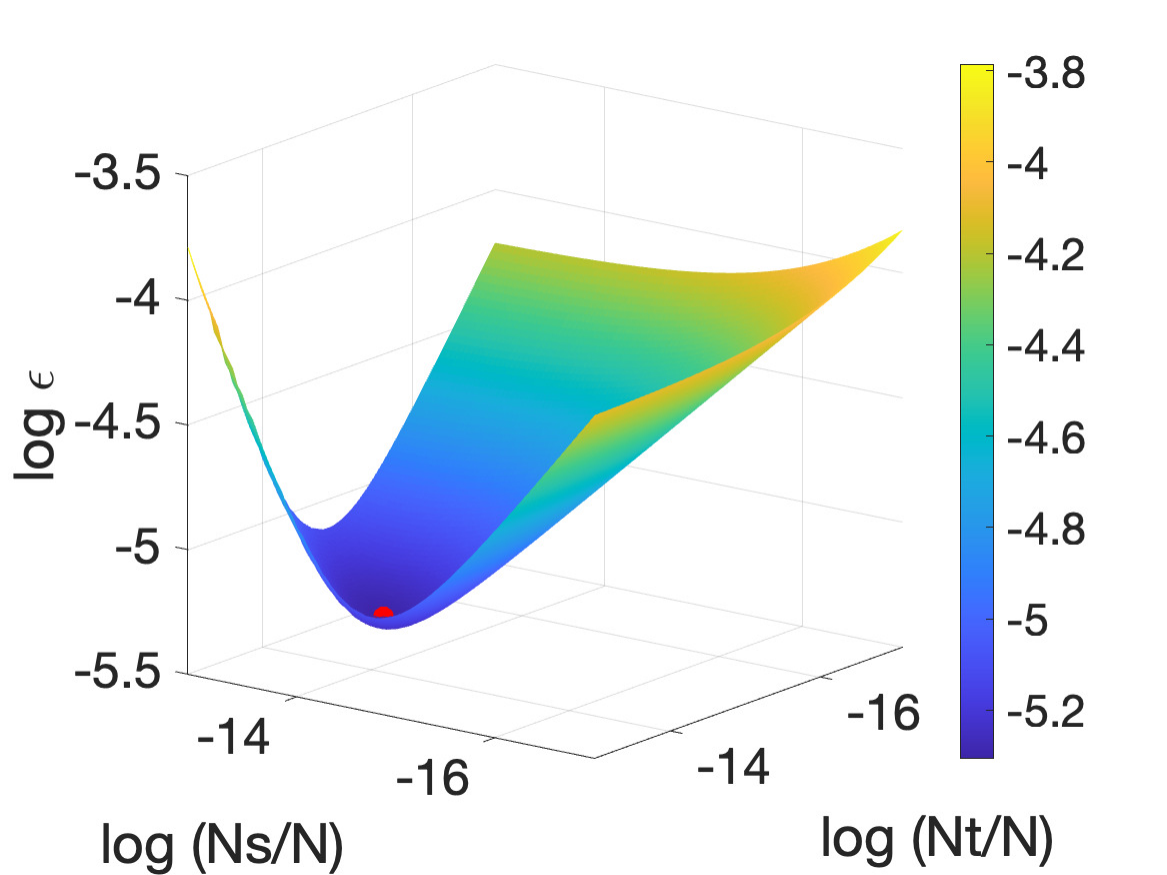}
\par\end{centering}
\caption{Total error $\epsilon$ versus the resources surface plot, for the
the midpoint (MP) method with an SPDE. $Ns/N$ is the ratio between
the sample size and the total resource, while $N_{t}/N$ is the ratio
between the number of time points and total resource. Here, $N$ is
$10^{10}$. The minimum error is indicated by the red circle. \label{fig:total_error_vs_resource}}
\end{figure}

With the estimated sampling, time-step and spatial-step error scalings,
the optimal resource for each error source $N_{i}$ is given by the
expression in Eq. (\ref{eq:complexity_prefactor})
\begin{align*}
\,\,N_{i}=\left(N^{c}\frac{\epsilon_{0i}n_{i}^{1/2}}{\epsilon_{0}c^{1/2}}\right)^{\frac{1}{n_{i}}} & \,,
\end{align*}
where $\epsilon_{0}=N_{A}^{c}c^{-1/p}\prod_{i}\left(n_{i}\epsilon_{0i}^{p}\right)^{c/pn_{i}}$
and $c=\left[\sum n_{i}^{-1}\right]^{-1}$ as in Eq. (\ref{eq:optimal complexity order}).
These optimal resources minimize the total error $\epsilon$ Eq. (\ref{eq:total error})
and we verify this by calculating the total error for different ranges
of resources and locating a minimum total error in the surface plot,
as shown in Fig. \ref{fig:total_error_vs_resource}. We chose the
total resource to be $10^{10}$, and both $N_{s}$ and $N_{t}$ ranged
from $4.1\times10^{2}$ to $2.5\times10^{4}$, while $N_{d}$ is fixed
by the constraint $N_{d}=N/(N_{s}N_{t})$. 

The surface plot in Fig. \ref{fig:total_error_vs_resource} has a
minimum total error at the predicted optimal resources in Eq. (\ref{eq:complexity_prefactor}),
which corresponds to $8006$ samples, 7440 time points and 167 spatial
points. The estimated parameters used are given in Table (\ref{tab:SPDE_error_scalings_estimations}).
We take $N_{A}=1$, since this does not change the optimum complexity,
but more generally the single use resource requirement $N_{A}$ should
be included.

Next, we estimate the complexity order of the MP method for this stochastic
partial differential equation. We compute the comparison error $\epsilon_{c}$
for a set of total resources $N$ ranging from $10^{7}$ to $10^{11}$.
For each value of total resource, we use the optimal resource for
each error source $N_{i}$, given by the expression in Eq. (\ref{eq:complexity_prefactor}). 

\begin{figure}[h]
\begin{centering}
\includegraphics[width=0.9\columnwidth]{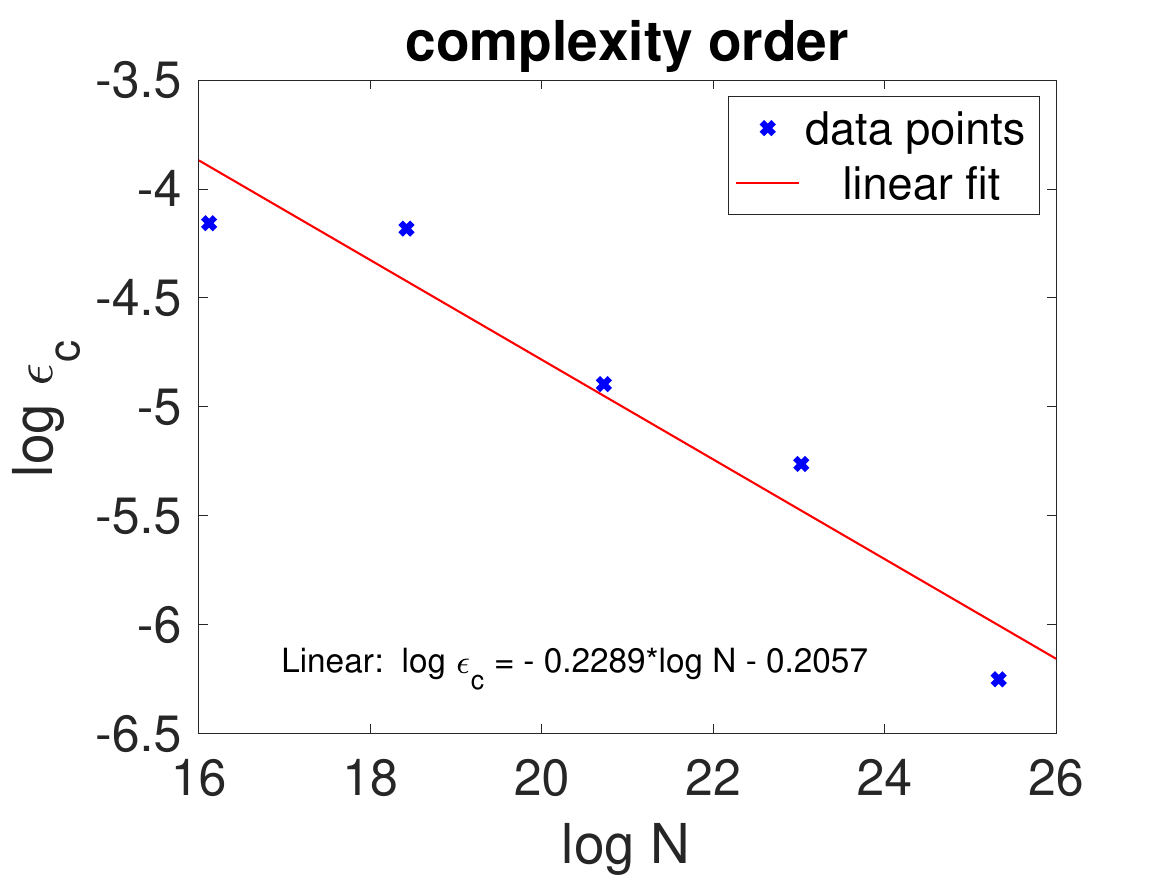}
\par\end{centering}
\caption{Complexity error scaling, for the the midpoint (MP) method with an
SPDE. The error scalings from the plot are tabulated in Table \ref{tab:SPDE_error_scalings_estimations}.
\label{fig:complexity_order_planar_noise}}
\end{figure}

The result for the complexity order estimation is presented in Fig.
\ref{fig:complexity_order_planar_noise}. The estimated complexity
order was $0.23\pm0.04$, which agrees with the theoretical prediction
of $0.2$ where the sampling and time-step error orders are $0.5$,
with a space-step order of $1$. The red line in the figure of the
best fit agrees reasonably well with the predicted optimal error from
Figure (\ref{fig:total_error_vs_resource}), which plots total error
vs resources.

\section{Conclusion}

The optimal complexity order of an algorithm with additive errors
and factorable resources is the inverse of the sum of each inverse
order. This is never better than its lowest order part. Thus, the
complexity order of SDE solvers with independent noise is never better
than $1/2$. Similar results hold for other algorithms with factorable
resources, including stochastic partial differential equation solvers.
This means that the advantage of higher-order solvers is less than
expected from the time-step order alone, and one must also include
the single step cost as a significant factor.

For resource usage in stochastic differential equations, an error
balance of $\epsilon_{S}\left(r\right)=\sqrt{2n}\epsilon_{T}\left(r\right)$
is optimal. Expending computational resources to reduce either the
sampling error or the step-size error below this optimum level is
not efficient. This is especially significant for large numbers of
variables, where computing a high-order step requires large resources.
This result may be improved through the use of more sophisticated
sampling methods \citep{morokoff1995quasi,caflisch1998monte,giles2008multilevel,giles2015multilevel,opanchuk2016parallel}.

Investigating the optimal resource allocation is important in large-scale
numerical modeling. This is an issue, for example, in climate studies
\citep{ishii2020d4pdf}, although such cases may not have factorable
resources, which would require a more sophisticated optimization.
In some convergence comparisons \citep{skeel1990method} for partial
differential equation algorithms, only convergence in space step is
studied. The complexity order for more closely coupled algorithms
is therefore an open problem, which we do not treat in detail here.

There is also a practical limitation. Our results focus on the asymptotic
complexity order. Yet numerical studies may not be in this limit.
When there are large deterministic parts to a stochastic equation,
the effective order at some finite time-step is not the asymptotic
order. Thus, the optimum resource ratio for a finite error may not
always be the asymptotic ratio. This depends on the details of the
problem, and we note that an empirical investigation of time-step
errors at different step-sizes may be required.

We have investigated our predictions numerically. Three different
SDE algorithms gave agreement with the complexity result for the exactly
soluble case of the Kubo oscillator. Extending this to an SPDE also
gave agreement in a three resource heat equation case, which is also
exactly soluble. Clearly, combining more errors and resources gives
lower complexity orders. Each resource requirement contributes errors.
Hence, one must use resources efficiently to minimize the error.

In summary, the complexity order of stochastic solvers is important,
owing to the widespread use of these methods. Here, we show that precision
improvements require an optimum allocation of resources. All the predicted
complexity orders have been verified numerically for both SDE and
SPDE cases. Similar criteria hold in other multiple resource cases.
In some cases, algorithms may not have the additivity that aids optimization
here. Our results are therefore an indication of more general multiple
resource allocation methods that may be required with different problems
and algorithms.
\begin{acknowledgments}
We thank M. D. Reid for useful discussions. We gratefully acknowledge
a grant from NTT Phi Laboratories. This publication was made possible
through the support of Grant 62843 from the John Templeton Foundation.
The opinions expressed in this publication are those of the author(s)
and do not necessarily reflect the views of the John Templeton Foundation.
\end{acknowledgments}

\section*{Appendix}

Numerical simulations were run using the Matlab package xSPDE4, available
on Github \citep{Drummond2024xSPDE4}, and checked with independent
codes. The inputs are a data structure (p), which is used to define
the parameters and functions: the derivative function (p.deriv), method
(p.method), observable (p.observe), and the comparison function (p.compare)
used to compute errors.

\subsection*{Numerical scripts for Kubo problems}

The xSPDE input script for the Kubo oscillator using the midpoint
(MP) method is given below. The inputs $ns$ and $nt$ are the number
of sub-ensembles and time steps, which define the resource usage.
The derivative function (p.deriv) combines drift and noise. The first
ensemble, p.ensembles(1), uses vector operations. The second ensemble,
p.ensembles(2), is computed in series, which was not used, while p.ensembles(3)
is the number of sub-ensembles computed in parallel.

{\ttfamily{}
\begin{lstlisting}[basicstyle={\ttfamily}]
function [e,data,p] = Kubo(ns,nt)
p.ensembles =  [2000,1,ns];   
p.initial   =  @(w,p) 1;
p.points    =  nt+1;           
p.ranges    =  5;              
p.checks    =  0;              
p.method    =  @MP;             
p.deriv     =  @(a,w,~) 1i*w.*a;
p.observe   =  @(a,~) real(a);  
p.compare   =  @(p) exp(-p.t/2);
[e,data,p]  =  xspde(p);        
end
\end{lstlisting}
}

The comparison function is used to compute the errors of the averages.
Error checking is turned off (p.checks=0) since the errors are calculated
from known results. The script for the Runge-Kutta (RK4) method is
similar except that the method is set to @RK4. {\ttfamily{}
\begin{lstlisting}[basicstyle={\ttfamily}]
function [e,data,p] = Kubo(ns,nt)
p.ensembles =  [2000,1,ns];   
p.initial   =  @(w,p) 1;
p.points    =  nt+1;           
p.ranges    =  5;              
p.checks    =  0;              
p.method    =  @RK4;             
p.deriv     =  @(a,w,~) 1i*w.*a;
p.observe   =  @(a,~) real(a);  
p.compare   =  @(p) exp(-p.t/2);
[e,data,p]  =  xspde(p);        
end
\end{lstlisting}
}

The script for the Kloeden-Platen weak second-order method (KPW2)
is given below:
\begin{lstlisting}[basicstyle={\ttfamily}]
function [e,data,p] = KuboKPW2(ns,nt)
p.ensembles = [2000,1,ns]; 
p.initial   =  @(w,p) 1;
p.points    =  nt+1;    
p.ranges    =  5;   
p.checks    =  0;      
p.method    =  @RKWP21; 
p.derivA    =  @(a,p)  -0.5*a;                
p.derivB    =  @(a,p)  1i*a;   
p.observe   =  @(a,~) real(a);
p.compare   =  @(p) exp(-p.t/2);
[e,data,p]  =  xspde(p); 
end
\end{lstlisting}
Here, the method is set to @RKWP21, and the time evolution equation
is an Ito SDE, as in Eq. (\ref{eq:Kubo_Ito}). This method has distinct
functions for the drift (p.derivA) and noise (p.derivB) coefficients,
which do not have internal noise arguments.

\subsection*{Numerical script for the stochastic heat equation}

In this case the space-time dimension is $p.dimensions=2$, a default
initial condition of $a=0$ is used, the extra input of $nd$ defines
the number of spatial steps, there are now two real noises per lattice
point, and the $p.linear$ function is added to specify the interaction
picture transforms.

{\ttfamily{}
\begin{lstlisting}[basicstyle={\ttfamily}]
function [e,data,p] = SPDE(ns,nt,nd)
p.dimensions = 2;
p.ranges    = [1,5];
p.points    = [nt+1,nd];
p.noises    = 2;
p.checks    = 0;
p.ensembles = [ns,1,10];
p.method    = @MP;
p.deriv     = @(a,w,p) (w(1,:,:)...
             +1i*w(2,:,:))/sqrt(2);
p.linear    = @(p) .5*p.Dx.^2; 
p.observe   = @(a,p) Int(a.*conj(a),p);
p.compare   = @(p) 5*sqrt(p.t/pi);
[e,data,p]  = xspde(p);
end
\end{lstlisting}
}

\subsection*{Curve-fitting}

Curve-fitting methods \citep{acton1966analysis} were computed with
the Matlab function \citep{MATLAB:2023} $fit$, and a fit type of
$poly1$. This assumes that the data is normally distributed, with
all variances the same, and with all probabilities derived from a
linear model. Each assumption is approximate, and so the error-bars
can underestimate the true errors.

The gradient $b$ and y-intercept $a$ of a linear fit $y=a+bx$ can
be expressed in terms of the variance of $x$ ($\sigma_{x}^{2}$),
variance of $y$ ($\sigma_{y}^{2}$), and the covariance $cov(x,y)$.
The exact formulae are $a=\bar{y}-b\bar{x}\,\;$ and $b=s_{xy}/s_{xx}$,
where 
\begin{align}
\bar{x} & =\frac{1}{n}\sum_{i=1}^{n}x_{i};\,\,s_{xx}=\sum_{i=1}^{n}\left(x_{i}-\bar{x}\right)^{2}\\
s_{yy} & =\sum_{i=1}^{n}\left(y_{i}-\bar{y}\right)^{2};\,\,s_{xy}=\sum_{i=1}^{n}\left(x_{i}-\bar{x}\right)\left(y_{i}-\bar{y}\right)\nonumber 
\end{align}

The error between the fitted point and data point $e_{i}\equiv y_{i}-\hat{y}_{i}=y_{i}-(a+bx_{i})$
has a variance of $s^{2}=\sum_{i=1}^{n}e_{i}^{2}/(n-2)$, giving the
standard errors for the gradient $b$ and y-intercept $a$ using:
\begin{equation}
\sigma(a)=s\sqrt{\frac{1}{n}+\frac{\bar{x}^{2}}{s_{xx}}},\,\,\sigma(b)=\frac{s}{\sqrt{s_{xx}}}\,.
\end{equation}

The error-bars generally agreed with the range of results obtained
when different random number seeds were used to generate independent
datasets. These were obtained from the Matlab ``rng'' function with
the ``shuffle'' setting, initializing random number seeds using
the system time.

\bibliographystyle{apsrev4-2}

\end{document}